\documentclass[aps,twocolumn,showpacs,nofootinbib]{revtex4}

\usepackage{amsmath}
\usepackage{amsfonts}
\usepackage{amssymb}
\usepackage{color}

\def\bc{\begin{center}}
\def\nno{\nonumber}
\def\ec{\end{center}}
\def\be{\begin{eqnarray}}
\def\ee{\end{eqnarray}}

\newcommand{\omits}[1]{}

\definecolor{dyellow}{rgb}{1.,0.8,.0}
\definecolor{myblue}{rgb}{.1,.1,.7}
\definecolor{dcyan}{rgb}{.0,.6,.6}
\definecolor{dmagenta}{rgb}{0.6,0.0,0.6}
\definecolor{brown}{rgb}{0.6,0.2,0.}
\definecolor{darkblue}{rgb}{.0,.0,0.5}
\definecolor{darkred}{rgb}{0.75,0.0,0.0}
\definecolor{orange}{rgb}{1.,.6,.0}
\definecolor{dorange}{rgb}{0.8,.4,.0}
\definecolor{darkgreen}{rgb}{0.0,0.6,0.0}
\definecolor{purple}{rgb}{.4,.0,.4}

\def\La{\Lambda}
\def\Si{\Sigma}

\def\dl{\delta}
\def\eps{\epsilon}

\def\la{\lambda}

\def\si{\sigma}

\def\d#1#2{\frac{\displaystyle #1}{\displaystyle #2}}

\newcommand{\vect}[1]{\mbox{\boldmath{$ #1$}}}
\newcommand\btd{\raise 2pt
\hbox{$\hat\bigtriangledown$}\hskip 1.5pt}
\newcommand\bt{\raise 2pt
\hbox{$\bigtriangledown$}\hskip 1.5pt}

\begin{document}



\title{On Special Relativity with Cosmological Constant}


\author{{Han-Ying Guo}$^{1,2}$}
\email{hyguo@itp.ac.cn}
\author{{Chao-Guang Huang}$^{3}$}
\email{huangcg@mail.ihep.ac.cn}
\author{{Zhan Xu}$^{4}$}
\email{zx-dmp@mail.tsinghua.edu.cn}
\author{{Bin Zhou}$^{2}$} \email{zhoub@itp.ac.cn}
\affiliation{%
${}^1$ CCAST (World Laboratory), P.O. Box 8730, Beijing
   100080, China,}

\affiliation{%
${}^2$ Institute of Theoretical Physics,
 Chinese Academy of Sciences,
 P.O.Box 2735, Beijing 100080, China,}

\affiliation{%
${}^3$ Institute of High Energy Physics, Chinese Academy of
Sciences, P.O. Box 918-4, Beijing
   100039, China,}
\affiliation{%
${}^4$ Physics Department, Tsinghua University, Beijing
   100084, China.}
\date{March 10, 2004}


\begin{abstract}
Based on  the principle of relativity and the postulate of
invariant speed and
length, we propose the theory of special relativity with
cosmological constant ${\cal SR}_{c,R}$ if the  invariant
length whose square is the inverse of the one-third cosmological
constant of the universe. It is on the Beltrami-de Sitter spacetime
${\cal B}_R$ with de Sitter invariance. We define the observables of free
particles and generalize famous Einstein's formula. We also define
two kinds of simultaneity. The first is for local experiments and
inertial motions. The second is for  cosmological
observations. Thus there is a relation between the relativity
principle and the cosmological principle. We predict that the 3-d
cosmic space is then  of positive spatial curvature
of order  cosmological constant.  The relation between ${\cal SR}_{c,R}$ and
the doubly special relativity is briefly disucssed.
\end{abstract}


\maketitle


\section{Introduction}
Recent observations show that there should be a positive
cosmological constant \cite{WMAP},\cite{SDSS}. These lead to a lot
of puzzles related to the de Sitter ($dS$) space 
and may touch
the foundation of physics \cite{Str}.

In this letter, we generalize Einstein's special relativity ${\cal
SR}_c$ to the one with an additional invariant constant of length,
${\cal SR}_{c, R}$, based on  the principle of relativity (PoR)
and the postulate of invariant speed and length (PoI). With an
identification of the length as $\sqrt{3/\La}$ where $\La$ is the
cosmological constant, it is the special relativity of the
Beltrami-de Sitter spacetime ${\cal B}_R$ with de Sitter
invariance \cite{BdS},\cite{NH}. It is of course
straightforward to set up its counterpart of anti-de Sitter(AdS)
spacetime with an identification of the length as $\sqrt{-
3/\La}$ ( For earlier works, see also \cite{Lu},\cite{LZG}).

It is related with some simple but important properties of $dS$
spacetime \cite{Gursey}--\cite{Guo1}. In fact, among various
metrics of $dS$ spacetime, there is an important one in which $dS$
spacetime is in analog with Minkowski spacetime. It is the $dS$
spacetime with Beltrami-like metric, called $BdS$ spacetime and
denoted as ${\cal B}_R$. $BdS$ spacetime is precisely the
Beltrami-like model \cite{beltrami} of a 4-hyperboloid ${\cal
S}_R$ in 5-d Minkowski spacetime, i.e. ${\cal B}_R \backsimeq
{\cal S}_R $. In ${\cal B}_R$ there exist a set of Beltrami
coordinate systems, which covers ${\cal B}_R$ patch by patch, and
in which test particles and light signals move along the timelike
and null geodesics, respectively, with {\it constant} coordinate
velocities. Therefore, they look like in free motion in a
spacetime without gravity. Thus, the Beltrami coordinates and
observers ${\cal O}_{B}$ at these systems should be regarded as of
inertia. And the classical observables for these particles and
signals can be well defined.

The key observation is that in a constant curvature spacetime
there do exist such motions and observers of inertia. If we start
with the 4-d Euclidean geometry and weaken the fifth axiom, then
there should be 4-d Riemann, Euclid, and Lobachevski geometries at
equal footing. Importantly, their geodesics are globally straight
lines in certain coordinate systems that are just ones in analog
with the coordinate systems in Beltrami model of Lobachevski plane
\cite{beltrami}, and under corresponding transformation groups the
systems transform among themselves. Now changing the signature to
$-2$, these constant curvature spaces turn to be $dS$, Minkowski($M$), and
$AdS$ spacetimes, respectively, and those straight
lines are classified by timelike, null and spacelike straight
world-lines. This may also be seen from viewpoint of projective
geometry by an antipodal identified version of, say,  $dS$ space
time, $dS/Z_2 \subset RP^4$, except  the latter %
is neither orientable nor time orientable.

Thus, in analog with ${\cal SR}_{c}$, all of them should describe
the inertial motions for free particles and light signals,
respectively. And, Einstein's PoR should also be available for
them, called the PoR with respect to $dS$, Poincar\'e, $AdS$
groups, respectively. In addition to the invariant speed, the
speed of light, in ${\cal SR}_c$ there is an invariant length $R$
in $dS$ and $AdS$, so the postulate of invariance of velocity of
light in ${\cal SR}_c$ should be replaced by the PoI. That is why
based on these two principles, the special relativity with
cosmological constant ${\cal SR}_{c,R}$ can be set up if the
length  being identified with $\sqrt{\pm 3/\La}$.

The Beltrami coordinates make sense and concern with the
measurements in laboratory at one patch, only if the simultaneity
should be defined with respect to the Beltrami time coordinate. On
the other hand, the simultaneity may also be defined by the proper
time of a clock rest at the origin of Beltrami spatial
coordinates. Thus, there are two kinds of simultaneity in ${\cal
SR}_{c,R}$ of ${\cal B}_R$. Consequently, if the simultaneity is
transformed from the first to the second, the Beltrami metric is
reduced to the Robertson-Walker(RW)-like metric in ${\cal B}_R$
with positive, rather than zero or negative, spatial curvature and
its deviation from zero is in the order of cosmological constant
$\Lambda$. Thus, the second simultaneity should be linked with the
observation in the cosmological scale. This important property
differs with either ${\cal SR}_{c}$ where two of them coincide, or
usual cosmological models where $k$ is a free parameter.  In fact,
this is an important prediction. The tiny spatial closeness seems
more or less to have been already confirmed by the CMB power
spectrum data from WMAP \cite{WMAP} with error-control from SDSS
\cite{SDSS} and should be further checked by its data in large
scale.

Finally, we briefly mention that one may regard ${\cal SR}_{c,R}$
of ${\cal B}_R$ as a counterpart of the ``doubly special
relativity"
(DSR) \cite{DSR} in its momentum space 
as long as $R$ is taken as an observer-independent large-momentum
scale.

This letter is organized as follows.  In section II, we present
two postulates and set up a framework for ${\cal SR}_{c,R}$ of
${\cal B}_R$. In section III we show explicitly the inertial
motion of particles and light signals, define their observables
and the simultaneity in inertial frame. In section IV, we define
another kind of simultaneity with proper time and show that ${\cal
B}_R$ can be reduced to RW-like metric with a slightly closed
space. We end with a few remarks including the relation between
the ${\cal SR}_{c,R}$ and the DSR.


\section{Relativity with Invariant Length}
We propose two fundamental principles as follows: The PoR requires
{\it there exist a set of inertial reference frames, in which the
free particles and light signals move with uniform velocities
along straight lines, the laws of nature without gravity are
invariant under the transformations among them.} The PoI
requires {\it there exist two invariant universal constants, one
is with dimension of velocity, and
another is of length.}

It can be proved that {\it the most general
form of the transformations among inertial coordinate systems
\be\label{FL}%
{x'}^i=f^i( x^j), \quad x^0=ct, \quad i,j=0,\cdots, 3,%
\ee
which transform a uniform-velocity, straight-line motion in $F$ to
a motion of the same nature in $F'$ are that the four functions
$f^i$ are ratios of linear functions, all with the same
denominators, where $c$ is the invariant speed} \cite{Fock}.

As in ${\cal SR}_c$, the PoR
implicates that if there exists a metric on inertial frame at
spacetime, it is of signature $\pm 2$ and invariant under some of these
transformations that form a group with ten parameters including
space-time ``translations" (4), boosts (3) and space rotations
(3), respectively. Thus, it can be proved that {\it
the necessary and sufficient condition for 4-d spacetimes with invariant metric of signature $\pm 2$ under
ten-parameter transformation group is that they are $dS/M/AdS$ of
positive, zero, or negative constant curvature, with the invariant
group $SO(1, 4)$, $ISO(1,3)$ or $SO(2, 3)$, respectively.} The PoI
requires that there is a room for the constant $c$ as in
(\ref{FL})  and the invariant length should be the
curvature radius of $dS/AdS$, respectively.

We now consider the case of $dS$ spacetime, which can then be
regarded as a 4-d hyperboloid ${\cal S}_R$ embedded in a 5-d
Minkowski spacetime with $\eta_{AB}= {\rm diag}(1, -1, -1, -1,
-1)$:
 \be\label{5sphr}%
 {\cal S}_R:  &&\eta^{}_{AB} \xi^A \xi^B= -R^2,%
\\ %
\label{ds2}%
&&ds^2=\eta^{}_{AB} d\xi^A d\xi^B , 
\ee
where $R$ is the invariant length, and $A, B=0, \ldots,
4$. Clearly, Eqs.(\ref{5sphr}) and (\ref{ds2}) are invariant under $dS$ group
${\cal G}_R =SO(1,4)$.

The Beltrami coordinates
are defined patch by patch on ${\cal B}_R\simeq{\cal S}_R$. For
intrinsic geometry of ${\cal B}_R$, there are at least eight
patches $U_{\pm\alpha}:= \{ \xi\in{\cal S}_R : \xi^\alpha\gtrless
0\}, \alpha=1, \cdots, 4$. In $U_{\pm 4}$, for instance, the
Beltrami coordinates are
\be \label{u4}%
&&x^i|_{U_{\pm 4}} =R {\xi^i}/{\xi^4},\qquad i=0,\cdots, 3;  \\
&&\xi^4|_{U_{\pm 4}}=({\xi ^0}^2-\sum _{a=1}^{3}{\xi ^a}^2+ R^2
)^{1/2} \gtrless 0.
\ee%
In the patches $\{U_{\pm a}, a=1,2,3\}$,
\begin{equation}  %
y^{j'}|_{U_{\pm a}}=R{\xi^{j'}} /{\xi^a},\quad
j'=0,\cdots,\hat{a}\cdots,4; \quad \xi^{a}\neq 0,
\end{equation}
where $\hat{a}$ means omission of $a$. It is important that all
transition functions in  intersections are of ${\cal G}_R$, say,
in $U_4\bigcap U_3$, the transition function $T_{4,3}
=\xi^3/\xi^4=x^3/R=R/y^4 \in {\cal G}_R$ so that
$x^i=T_{4,3}y^{i'} $  for $i=i'=0,1,2$ and $x^3=R^2/y^4$. Note all
 of them are of type (\ref{FL}).

It is important that {\it in each patch, there are condition and
Beltrami metric
\begin{eqnarray}\label{domain}
\sigma(x)&=&\sigma(x,x):=1-R^{-2}
\eta_{ij}x^i x^j>0,\\
\label{bhl} ds^2&=&[\eta_{ij}\sigma(x)^{-1}+ R^{-2}
\eta_{ik}\eta_{jl}x^k x^l \sigma(x)^{-2}]dx^i dx^j.
\end{eqnarray}
 Under  fractional linear
transformations of type (\ref{FL})
\begin{equation}\label{G}
\begin{array}{l}
x^i\rightarrow \tilde{x}^i=\pm\sigma(a)^{1/2}\sigma(a,x)^{-1}(x^j-a^j)D_j^i,\\
D_j^i=L_j^i+{ R^{-2}}%
\eta_{jk}a^k a^l (\sigma(a)+\sigma(a)^{1/2})^{-1}L_l^i,\\
L:=(L_j^i)_{i,j=0,\cdots,3}\in SO(1,3),
\end{array}\end{equation}
 where $\eta_{ij}%
={\rm diag} (1, -1,-1,-1)$ in $U_{\pm\alpha}$, Eqs. (\ref{domain})
and (\ref{bhl}) are invariant.  
Here (\ref{G}) are the transitive part to the origin of
${\cal G}_R$. The inertial frames and inertial motions transform
among themselves, respectively.}

Note that Eqs. (\ref{domain})-(\ref{G}) are defined on ${\cal
B}_R$ patch by patch. $\sigma(x)=0$ is
the conformal boundary of ${\cal B}_R$, $\partial{\cal B}_R$.

The generators of ${\cal G}_\Lambda$ in Beltrami coordinates are expressed as %
\begin{equation}\label{generator}
\begin{array}{l}
  \mathbf{P}_i =(\delta_i^j-R^{-2}x_i x^j) \partial_j, \quad  x_i:=\eta_{ij}x^j,\\
  \mathbf{L}_{ij} = x_i \mathbf{P}_j - x_j \mathbf{P}_i
  = x_i \partial_j - x_j \partial_i \in so(1,3).
\end{array}
\end{equation}
They form an $so(1,4)$ algebra
\begin{eqnarray}
  [ \mathbf{P}_i, \mathbf{P}_j ] &=& R^{-2} \mathbf{L}_{ij} \nno\\
  {[} \mathbf{L}_{ij},\mathbf{P}_k {]} &=&
    \eta_{jk} \mathbf{P}_i - \eta_{ik} \mathbf{P}_j
\label{so(1,4)}\\
  {[} \mathbf{L}_{ij},\mathbf{L}_{kl} {]} &=&
    \eta_{jk} \mathbf{L}_{il}
  - \eta_{jl} \mathbf{L}_{ik}
  + \eta_{il} \mathbf{L}_{jk}
  - \eta_{ik} \mathbf{L}_{jl}. \nno
\end{eqnarray}

For two separate events $A(a^i)$ and $X(x^i)$ in ${\cal B}_R$,
\be\label{lcone0} %
{\Delta}_R^2(A, X) = R
[\sigma^{-1}(a)\sigma^{-1}(x)\sigma^2(a,x)-1]
\ee %
is invariant under transformations (\ref{G}) of ${\cal G}_R$.
Thus, the interval between $A$ and $B$ is timelike, null, or
spacelike,
respectively, according to%
\begin{equation}\label{lcone}%
\Delta_R^2(A, B)\gtreqqless 0.%
\end{equation}

The proper length of timelike or spacelike between $A$ and $B$ are
integral of ${\cal I} ds$ over the geodesic segment
$\overline{AB}$:
\be \label{AB1}%
S_{t-like}(A,B)&=&R \sinh^{-1} (|\Delta(a,b)|/R), \\
\label{AB1sl} S_{s-like}(A,B)&=& R \arcsin (|\Delta(a,b)|/R),
\ee
where ${\cal I}=1, -i$ for timelike or spacelike,
respectively.

It can be shown that the light-cone at $A$ with running points
$X$ is
\be \label{nullcone} %
{\cal F}_{R}:= R
\{\sigma(a,x) \mp [\sigma(a)\sigma(x)]^{1/2}\}=0.%
 \ee%
It satisfies the null-hypersurface condition
 \be\label{Heqn}%
\left . g^{ij}\frac{\partial f}{\partial x^i}\frac{\partial
f}{\partial x^j}\right |_{f=0}=0, %
\ee
where $g^{ij}=\sigma(x)(\eta^{ij}-R^{-2} x^i x^j)$ is the inverse
Beltrami metric. And at the origin of the Beltrami coordinates
$a^i=0$, the light cone becomes a Minkowskian one and the
invariant speed $c$ is numerically the velocity of light in the
vacuum.


\section{Inertial Motion, Observable and Beltrami Simultaneity}

According to the PoR, a free particle with mass $m_{\Lambda 0}$
should move inertially along a timelike geodesic of the $BdS$
spacetime. It is just the case. The geodesic equation has the
first integration
\begin{equation}\label{pi}
\frac{dp^i}{ds}=0, \quad
p^i:=m_{\Lambda 0}\sigma(x)^{-1}\frac{dx^i}{ds}=C^i={\rm const.}
\end{equation}
This implies under the initial condition
\[
x^i(s=0)=b^i, \qquad \frac{dx^i}{ds}(s=0)=c^i
\]
with the constraint
$
g_{ij}(b)c^i c^j=1,
$
a new parameter $w = w(s)$ can be chosen such that
the geodesic is just a straight world-line
\begin{equation}
x^i(w)=c^iw+b^i.
\end{equation}
The parameter $w$ can be integrated out,
\be \label{w1}
w(s) = \begin{cases}
R e^{\mp s/R}\sinh \d s R,   &    \eta_{ij}\,c^i c^j = 0, \medskip \cr
\d {R\sinh \frac s R} {\frac {\eta_{ij}\,c^i b^j}{R\sigma(b)}\sinh \frac s R
    + \cosh \frac s R},  & \eta_{ij}\,c^i c^j \neq  0.
\end{cases}
\ee

Similarly, a light signal moves inertially along a null geodesic, which
still has the first integration
\begin{equation}
\sigma ^{-1}(x)\frac{dx^i}{d\tau }={\rm constant},
\end{equation}
where $\tau $ is an affine parameter.  Again, under the
initial condition
\be %
x^i(\tau =0)=b^i,  \qquad \d {dx^i}{d\tau }(\tau =0)=c^i. %
\ee
and the constraint $ g_{ij}(b)\,c^ic^j=0$, the null geodesic can
be expressed as a straight line
\be x^i = c^i w(\tau) +b^i, \ee
where
\be \label{w2}%
w(\tau) = \begin{cases} {~ \tau}, & \eta _{ij}\,c^ic^j =0, \cr -
\d { R^2\sigma (b)}{|\eta _{ij}c^ic^j|}\left( \d 1{\tau +\tau
_0}-\d 1{\tau _0}\right), & \eta _{ij}\,c^ic^j \neq 0,
\end{cases}
\ee
with
\begin{equation}\nonumber
\tau _0=\sqrt{\frac{ R^2 \sigma (b)}{|\eta _{ij}c^ic^j|}}.
\end{equation}

Thus, the  motions of both free particles and light signals are
indeed inertial, i.e. the coordinate velocity
components  are constants, respectively:
\begin{equation}\label{vi}
\frac{dx^a}{dt}=v^a;\quad \frac{d^2x^a}{dt^2}=0;\quad a=1,2,3.
\end{equation}
Note that these properties are well defined patch by patch on whole
${\cal B}_R$.

Now we are ready to define the observables for free particles.
From  Eq. (\ref{pi}), it is natural to define the conservative
quantities $p^i$
as the 4-momentum of a free particle with mass $m_{\Lambda, 0}$
and its zeroth component as the energy. Note that it is no
longer a 4-vector rather a pseudo 4-vector.

Furthermore, for a free particle a set of conserved quantities
$L^{ij}$ may also be defined by
\begin{equation}\label{angular4}
L^{ij}=x^ip^j-x^jp^i;\quad \frac{dL^{ij}}{ds}=0.
\end{equation}
These may be called the 4-angular-momentum and they
are also no longer an anti-symmetric tensor but a pseudo anti-symmetric tensor.
However, $p^i$ and $L^{ij}$ constitute a conserved 5-d angular
momentum for a free  particle  such that ${\cal L}^{ij} = L^{ij}$,
${\cal L}^{4i} = p^i$.  Then, we have
\begin{equation}\label{angular5}
{\cal L}^{AB}=m_{\Lambda
0}(\xi^A\frac{d\xi^B}{ds}-\xi^B\frac{d\xi^A}{ds}); \quad
\frac{d{\cal L}^{AB}}{ds}=0.
\end{equation}
For such a kind of free particles, there is a {\it
generalized Einstein's famous formula}  in ${\cal B}_R$:
\begin{equation}\label{eml}
-\frac{1}{2R^2}{\cal L}^{AB}{\cal L}_{AB}=E^2-\vect{P}\,^2-
\frac{1}{2R^2} L^{ij}L_{ij}=m_{\Lambda 0}^2,
\end{equation}
where ${\cal L}_{AB}=\eta_{AC}\eta_{BD}{\cal L}^{CD}$,
$L_{ij}=\eta_{ik}\eta_{jl}L^{kl}$, italic blod $\vect{P}$ denotes the triple $(P^1,P^2,P^3)$, $m_{\Lambda 0}$ %
introduced above should be the inertial mass for a free particle.
It is well defined together with the energy, momentum and angular
momentum at classical level.

It can further be shown that $m_{\Lambda 0}^2$ is the eigenvalue of
the first Casimir operator of 
${\cal G}_R$ \cite{Gursey}
\be \label{C1}
\mathbf{C}_1 = \mathbf{P}_i \mathbf{P}^i -\frac 1 2 R^{-2} \mathbf{L}_{ij} \mathbf{L}^{ij}
\ee
with $\mathbf{P}^i = \eta^{ij} \mathbf{P}_j, \mathbf{L}^{ij}=\eta^{ik} \eta^{jl} \mathbf{L}_{kl}$.  In addition, spin can also be defined through the
eigenvalue of the second Casimir orperator
\be
\mathbf{C}_2 = \mathbf{S}_i \mathbf{S}^i -R^{-2} \mathbf{W}^2
\ee
as it was done in the relativistic quantum mechanics in Minkowski spacetime,
where%
\be \label{sw} %
\begin{array}{l}
\mathbf{S}_i = \frac 1 2 \eps_{ijkl} \mathbf{P}^j \mathbf{L}^{kl}, \quad %
\mathbf{S}^i =\eta^{ij} \mathbf{S}_j,\\
\mathbf{W} =\frac 1 8 \eps_{ijkl} \mathbf{L}^{ij} \mathbf{L}^{kl}. %
\end{array}
\ee
In Eq.(\ref{sw}), $\eps_{\mu \nu \la \si}$ is 4-d Levi-Civita
symbol in flat spacetime with $\eps_{0123}=1$.

Thus, ${\cal SR}_{c, R}$ offers a consistent way to define the
observable for free particles.  Of course, these issues
significantly confirm  that the motion of a free particle in
${\cal B}_R$ is of inertia, the coordinate systems with Beltrami
metric are  inertial systems and corresponding observer at the
origin of the system is of inertia as well.

Note that in order to make sense of inertial motions, etc., one
should define simultaneity and take space-time measurements.
In ${\cal SR}_{c}$, coordinates have measurement significance
that is linked with the PoR. Namely, the difference in time
coordinate stands for the time interval, and the difference in
spatial coordinate stands for the spatial distance. Similar to
${\cal SR}_{c}$, one can define that two events $A$ and $B$ are
simultaneous if and only if the Beltrami time coordinate $x^0$ for
the two events are same, 
\be %
a^0:=x^0(A) =x^0(B)=:b^0. %
\ee
It is called the {\it Beltrami simultaneity} and with respect to
it that free particles move inertially.
The Beltrami simultaneity defines a 3+1 decomposition of spacetime
\be ds^2 =  N^2 (dx^0)^2 - h_{ab} \left (dx^a+N^a dx^0 \right )
\left (dx^b+N^b dx^0 \right ) %
\ee
with the lapse function, shift vector, and induced 3-geometry on
3-hypersurface $\Si_c$ in one coordinate patch.
\begin{eqnarray}
& & N=\{\si_{\Si_c}(x)[1-(x^0 /R)^2]\}^{-1/2}, \nonumber \\%
& & N^a=x^0 x^a[ R^2-(x^0)^2]^{-1},
 \\
& & h_{ab}=\dl_{ab} \si_{\Si_c}^{-1}(x)-{ [R\si_{\Si_c}(x)]^{-2}
\dl_{ac} \dl_{bd}}x^c x^d ,\nonumber
\end{eqnarray}
respectively, where $\si_{\Si_c}(x)=1-(x^0{/R})^2 + {\dl_{ab}x^a
x^b /R^2}$,  $\dl_{ab}$ is the Kronecker $\dl$-symbol,
$a,b=1,2,3$.  In particular, at $x^0=0$, $\si_{\Si_c}(x)=1+{
\dl_{ab} x^a x^b/R^2}$. In a vicinity of the origin of Beltrami
coordinate system in one patch, 3-hypersurface $\Si_c$ acts as a
Cauchy surface.

The Beltrami simultaneity defines the laboratory time in one
patch. According to the spirit of ${\cal SR}_{c}$ as well as the
PoR, the Beltrami coordinates define, in such a manner, the
standard clocks and standard rulers in laboratory of ${\cal
SR}_{c,R}$ on ${\cal B}_R$. To measure the time of a process or
the size of an object, we just need to compare with Beltrami
coordinates.

\section{Proper-time Simultaneity and Its Cosmological Significance}

There is another simultaneity, however, in ${\cal SR}_{c, R}$. It
is {\it proper-time simultaneity} with respect to
a clock rest at spatial origin of
the Beltrami coordinate system. Such a proper time
$\tau_{\Lambda>0}$ of a rest clock on the time axis of Beltrami
coordinate system, $\{x^a=0\}$, reads
\begin{eqnarray}\label{ptime}
\tau_{\Lambda>0}=R \sinh^{-1} (R^{-1}\sigma^{-\frac{1}{2}}(x)x^0).
\end{eqnarray}%
Therefore, we can define that the events are simultaneous with
respect to it if and only if
\begin{equation}\label{smlt}
x^0\sigma^{-1/2}(x,x)=\xi^0:=R \sinh(R^{-1}\tau)=\rm constant.
\end{equation}
 The line-element %
on the simultaneous 3-d
hypersurface, denoted by ${\Sigma_\tau}$, can be defined as
\begin{equation}\label{dl}
dl^2=-ds^2_{\Sigma_\tau},%
\end{equation}
where
\be \begin{array}{l}\label{spacelike}
ds^2_{\Sigma_\tau} = R_{\Si_\tau}^2%
dl_{{\Sigma_\tau} 0}^2, \\
R_{\Sigma_\tau}^2%
:=\sigma^{-1}(x,x)\sigma_{\Sigma_\tau}(x,x)%
=1+ (\xi^0/R)^2,\\%
\sigma_{\Sigma_\tau}(x,x):=1+R^{-2}\delta_{ab}x^a x^b
>0, \\ 
dl_{{\Sigma_\tau} 0}^2:={
\{\delta_{ab}\sigma_{\Sigma_\tau}^{-1}(x)
-[R\sigma_{\Sigma_\tau}(x)]^{-2}\delta_{ac}\delta_{bd}x^c x^d\}}
 dx^a dx^b.
\end{array}\ee

It should be pointed out that this simultaneity is closely linked
with the cosmological principle. In fact,
 it is significant that if $\tau_{\Lambda>0}$ is taken as
a ``cosmic time", the Beltrami metric (\ref{bhl}) becomes an
RW-like metric with a positive spatial curvature and the
simultaneity is globally defined in whole ${\cal B}_R$
\begin{equation}\label{dsRW}
ds^2=d\tau^2-dl^2=d\tau^2-\cosh^2( R^{-1}\tau) dl_{{\Sigma_\tau} 0}^2.%
\end{equation}
This  shows that {\it the 3-d cosmic space is $S^3$ rather than
flat. The deviation from the flatness is of order $\Lambda$.}
Obviously, this spatial closeness of the universe is a remarkable
property different from the standard cosmological model with
flatness. This property seems more or less already indicated by
the CMB power spectrum from WMAP \cite{WMAP},\cite{SDSS} and
should be further checked by its data in large scale.

The two definitions of simultaneity do make sense in different
kinds of measurements. The first concerns the measurements in a
laboratory and is related to the PoR of
${\cal SR}_{c, R}$. The second concerns with cosmological effects.
Furthermore, the relation between the Beltrami metric
and its RW-like counterpart (\ref{dsRW})
in terms of the coordinate time $x^0$ and the cosmic time
$\tau$ links the PoR and cosmological principle. It is very meaningful.


\section{Remarks}

We have set up the special relativity with an invariant length $R$
in addition to $c$, ${\cal SR}_{c, R}$ of ${\cal B}_R$, with $dS$
invariance based on the PoR and PoI. Similarly, ${\cal SR}_{c, R}$
with $AdS$ invariance can also be set up. The Beltrami coordinates
are of inertia, the test particles and signals move inertially
along the timelike, null straight world lines, respectively. Their
classical observable can be well defined. And, the famous
Einstein's energy-momentum-mass relation is generalized in ${\cal
SR}_{c,R}$.

The relation between the Beltrami metric and the
RW-like metric in ${\cal SR}_{c,R}$ of
${\cal B}_R$ indicates a relation between the PoR and
cosmological principle and predicts that the 3-d cosmic space is slightly closed.

It should be noted that all properties in ${\cal SR}_{c, R}$ of
${\cal B}_R$ are in analog with ${\cal SR}_{c}$  and coincide with
it if $R\to \infty$.

In fact, all possible kinematics with ten-parameter transformation
groups \cite{Bacry} should be contracted \cite{IW} from the
relativity with invariant speed and length ${\cal SR}_{c,R}$.  For
instance, one may consider the Newton-Hooke limit
\cite{ABCP},\cite{Gao},\cite{Gibbons} of the $dS$ spacetime.
However, as was mentioned in \cite{NH}, the Newton-Hooke limit is
{\it Not} unique. In order to fix the Newton-Hooke limit uniquely,
one should require the Galilei-Hooke's relativity principle as
well as the principle of infinite velocity of signal, which are
corresponding to the PoR and the PoI here, respectively.  Thus,
the Newton-Hooke space-time given in \cite{NH} is the Newton-Hooke
contraction of the relativity ${\cal SR}_{c,R}$ on ${\cal B}_R$.

Finally, we should mention the relation between the relativity
with invariant speed and length, ${\cal SR}_{c,R}$, presented here
and the DSR \cite{DSR}. In fact, ${\cal SR}_{c,R}$ is a relativity
with the both observer-independent invariants, velocity scale and
length scale, in configuration space, while the DSR is the one
involves an observer-independent large-velocity scale and an
observer-independent  small-length scale/large-momentum scale. It
has been noticed, however, that the DSR may be viewed as a theory
with energy-momentum space being the 4-d $dS$ space and that 
the inhomogeneous projective coordinates are  employed
\cite{dSDSR}, which are locally the same as the Beltrami
coordinates except the latter without the antipodal
identification. This may be traced back to \cite{Snyder}. However,
since the antipodal identification had been used there, resulting
in the energy-momentum space neither orientable nor energy
orientable. Thus, we may directly regard ${\cal SR}_{c,R}$ as the
counterpart of DSR in its momentum space without orientation
problem as long as $R$ is taken as the observer-independent
large-momentum scale.



\begin{acknowledgments}
The authors would like to thank Professors Z. Chang, %
G. Gibbons, Q. K. Lu, J. Z. Pan and Y.S. Wu as well as Drs. Y.
Ling and Y. Tian for valuable discussions. This work is partly
supported by NSFC under Grants Nos. 90103004, 10175070, 10375087.
\end{acknowledgments}

\end{document}